\documentclass[preprintnumbers,amsmath,11pt,amssymb,floatfix,superscriptaddress,nofootinbib]{article}

\def\mytitle#1{\setcounter{equation}{0}
\setcounter{footnote}{0}
\begin{flushleft}\Large\textbf{#1}\end{flushleft}
\vspace{0.25cm}}
\def\myname#1{\leftline{{\large #1}}\vspace{-0.13cm}}
\def\myplace#1#2{\small\begin{flushleft}\textit{#1}\\
\texttt{#2}\end{flushleft}}

\def\myclassification#1{\small\noindent
Keywords :
       #1\vspace{0.5cm}}
\usepackage{graphicx}
\usepackage{amsmath}
\begin{document}
\mytitle{Thermodynamic Variables of First-Order Entropy Corrected Lovelock-AdS Black Holes : P-V Criticality Analysis}

\myname{$Amritendu~ Haldar^{*}$\footnote{amritendu.h@gmail.com} and $Ritabrata~
Biswas^{\dag}$\footnote{biswas.ritabrata@gmail.com}}
\myplace{*Department of Physics, Sripat Singh College, Jiaganj, Murshidabad $-$ 742123, India.\\$\dag$ Department of Mathematics, The University of Burdwan, Golapbag Academic Complex, City : Burdwan  $-$ 713 104, District : Purba Barddhaman, State : WestBengal, India.} {}
\begin{abstract}
We investigate the effect of thermal fluctuations on the thermodynamics of a Lovelock-AdS black hole. Taking the first order logarithmic correction term in entropy we analyze the thermodynamic potentials like Helmholtz free energy, enthalpy and Gibbs free energy. We find that all the thermodynamic potentials are decreasing functions of correction coefficient $\alpha$. We also examined this correction coefficien tmust be positive by analysisng $P-V$ diagram. Further we study the P-V criticality and stability and find that presence of logarithmic correction in it is necessary to have critical points and stable phases. When P-V criticality appears, we calculate the critical volume $V_c$, critical pressure $P_c$ and critical temperature $T_c$ using different equations and show that there is no critical point for this black hole without thermal fluctuations. We also study the geometrothermodynamics of this kind of black holes. The Ricci scalar of the Ruppeiner metric is graphically analysed.
\end{abstract}

\myclassification{Thermal fluctuations; logarithmic correction; Thermodynamic potentials; P-V criticality.}

\section{Introduction:}
Black holes (BHs hereafter) are end state of massive stars which do not even let light to escape from them. General relativistically they can be understood as a singularity at $r=0$ which is wrapped by another singularity named as event horizon, crossing which from outside to inside, the role of time and space coordinates swap each other's role. After the proposition of Hawking's radiation \cite{Hawking 1975} which provides a real connection among the gravity and the quantum mechanics, a lot of attentions have been attracted to study the thermodynamic properties of the BHs, more specially when Hawking and Page completed their seminal paper \cite{Hawking 1983} which proposes that there may exist a phase transition in the phase space between Schwarzschild-AdS (Anti-de Sitter) BH and thermal radiation. This Hawking-Page phase transition can be explained as the confinement/deconfinement phase transition of gauge field in the AdS/CFT (Conformal Field Theory) correspondence \cite{Witten 1998}. The thermodynamic properties of BHs in AdS space are not exactly same for those BHs in de Sitter space or asymptotically flat spacetime. In AdS space, large BHs are thermodynamically stable and have positive specific heats, get cooler due to its loss of mass. On the other hand, the small BHs are thermodynamically unstable and have negative specific heat, get hotter due to their negative specific heats and eventually evaporate\cite{biswas1, biswas2, biswas3, biswas3, biswas4, biswas5, biswas6, biswas7, Choudhury1, Haldar 2018}.     

The original entropy of a BH is given by $S_0=A/4$, where $A$ is the area of event horizon of the BH in AdS space and the corrected entropy may be written as $S=S_0+ \alpha ln A+ \gamma_1 A^{-1}+ \gamma_2 A^{-2}+........ $, where $\alpha, \gamma_1, \gamma_2,......$ are the coefficients which depend on different BH thermodynamic parameters \cite{Upadhyay 2017}. The holographic principle which equates the degrees of freedom (DOF hereafter) of the boundary to any other region of space is inspired by aforementioned entropy equation and the modified entropy is due to quantum gravity corrections. The holographic principle will be corrected near the Planck scale and the quantum gravity corrections modify the topology of space-time at this scale. The corrected thermodynamics of some kinds of BHs such as G$\ddot{o}$del like BH \cite{Pourdarvish 2013} has heen studied by using the modified form of entropy $S=S_0+ \alpha ln A$. Then one should study the corrected thermodynamics of BHs by using the non-perturbative quantum gravity. The authors in \cite{Govindarajan 2001} have studied the effects of quantum correction to the BH thermodynamics with the help of Cardy formula. The corrected thermodynamics of BHs has been studied under the effect of matter field around the BHs \cite{Mann 1998, Medved 1999, Medved 2001}. The authors in \cite{Solodukhin 1998, Sen 2011, Sen 2013, Lowe 2010} have also studied the thermodynamic corrections produced by string theory which are in agreement with the other approaches to quantum gravity. The corrected thermodynamics of a dilatonic BH has also been discussed and observed to have the same universal manner \cite{Jing 2001}. The partition function of a BH plays an important role in studying the corrected thermodynamics of a BH \cite{Birmingham 2001}. The generalized uncertainty principle which helps us to yield the logarithmic correction \cite{Ali 2012, Faizal1 2015} is also useful to study the corrected thermodynamics of a BH in agreement with the other approaches to quantum gravity. The thermal fluctuations in the  BH thermodynamics would be obtained from a quantum correction to the space-time topology and it is of the same form that is as expected from the quantum gravitational effects \cite{Das 2002,  More 2005, Sadeghi 2014}. 

There are several papers in which the quantum corrections have already been used to study the BH geometries. The authors in \cite{Pourhassan 2015} have studied the logarithamic correction      
of the entropy of an AdS charged BH and found that the thermodynamics of the AdS BH is modified due to the thermal fluctuations. When the effects of thermal fluctuations on the thermodynamics of a modified Hayward BH have been studied, it has been found that the thermal fluctuations reduce the internal energy and pressure of that BH \cite{Pourhassan1 2016}. The thermal fluctuations for a black saturn \cite{Faizal2 2015} as well as a charged dilatonic black saturn \cite{Pourhassan2 2016} have been investigated. For the black saturn, it has been found that the thermal fluctuations do not have the major effects on stability of it. Again for a charged dilatonic black saturn when the thermal fluctuations have been studied either using the fluctuations in the energy or using the conformal field theory the model produces the same results. Though this result may differ for the other BHs. The logarithmic correction in entropy plays an important role for a sufficiently small BH and that is concluded after investigation of the thermodynamic properties of a small spinning Kerr- AdS BH under the effects of thermal fluctuations \cite{Pourhassan3 2016}. In \cite{Sadeghi 2016}, the authors have studied the logarithmically corrected thermodynamics of a dyonic charged AdS BH which is holographic dual of a Van der Waals' fluid is considered with the results that holographic picture is valid.         

While BH thermodynamics is considered, natures of phase transitions can be best investigated by the concept of geometry of thermodynamics. The Thermodynamic line element should be constructed and curvature can be interpreted. The curvature singularities of the thermodynamic metric is studied in many literatures. For AdS space it is studied in \cite{Quevedo2008}. The introduction of a Riemannian metric in thermodynamic system was firstly introduced by Weinhold \cite{Weinhold1, Weinhold2} as
\begin{equation}\label{ah3_equn_1}
g^W=\frac{\partial^2 M}{\partial X^i\partial X^j}dX^idX^j,~~ X^i=X^i(S,Q)
\end{equation}
where $S$ is the entropy and $Q$ is charge or any other thermodynamic parameter. Another metric was defined by Ruppeiner as \cite{Ruppeiner1, Ruppeiner2}
\begin{equation}\label{ah3_equn_2}
g^W=\frac{\partial^2 S}{\partial Y^i\partial Y^j}dY^idY^j,~~ Y^i=Y^i(M,Q)~~.
\end{equation}
These two metrics are related as
\begin{equation}\label{ah3_equn_3} 
ds^2_{R}=\frac{1}{T}ds^2_{W}~~.
\end{equation}  
Divergence in the Ricci scalar of Ruppeiner metric can speculate about the phase transitions.

The effect of quantum correction on the black hole thermodynamics for the LMP solution of Horava Lifshitz black hole in flat, spherical and hyperbolic spaces has been studied in \cite{Pourhassan1}. Quantum corrections to thermodynamics of quasitopological black holes is studied by Upadhyay, S. \cite{Upadhyay1}. This work has found a critical horizon radius for total mass density. Thermodynamics of a black hole geometry with hyperscaling violation is studied in \cite{Pourhassan2}. Effect of thermal fluctuations on the thermodynamics of massive gravity black holes in four dimensional AdS space is studied in \cite{Upadhyay2}. High curvature BTZ black holes' heat capacity and free energy and the geometric thermodynamics is studied in \cite{Hendi1}. Thermodynamics of higher order entropy corrected Schwarzschild Beltrami-deSitter Black Hole is studied in \cite{Pourhassan3}. Stability of Van der Waal's black holes in presence of logarithmic correction is studied in \cite{Upadhyay3}. Thermodynamic geometry of a static black hole in $f(R)$ gravity is studied in \cite{Upadhyay4}.

Our objective in this paper is to study the effect of thermal fluctuations on the thermodynamic properties of a Lovelock-AdS BHs. Taking the first order logarithmic correction term in entropy we obtain the thermodynamic potentials like Helmholtz free energy, enthalpy and Gibbs free energy of the said BH and we observe that all the thermodynamic potentials are decreasing functions of correction coefficient $\alpha$. We analyze the Helmholtz free energy $F$ relative to correction coefficient $\alpha$ and we notice that for higher values of AdS radius $R$ the values of critical radius also higher, though its variation very small relative to $R$. We compute the P-V behavior of the Lovelock-AdS BH in order to steady the effect of thermal fluctuations. We wish to study the geometrothermodynamics for this type of black holes as well. 

This paper is organized as follows: in the next section, we study the thermodynamic properties of Lovelock-AdS BHs. Here we introduce the first order correction to entropy (i.e., the logarithmic corrected entropy) as leading order of thermal fluctuation. In section 3, we analyze the $P-V$ criticality and stability of the BH. In section 4, we study the geometrothermodynamics for the thermodynamic system. Finally, in the last section, we present conclusion of the work.            
\section{Thermodynamics of Lovelock-AdS Black Holes:}

The Lovelock model \cite{Lovelock 1971} is considered as the most natural generalization of general relativity. One of the most important property of Lovelock Lagrangian is that though it may contain higher order terms itself , it yields field equations which are in second order in the metric. Also this theory gives the solutions which are free from ghosts. The action for asymptotically AdS space-time in the Lovelock model may be written in terms of Riemann curvature (given as $R^{ab}=d\omega^{ab}+\omega_c^a\omega^{cb}$) and the vielbein ( noted as $e^a$) as:

\begin{equation}\label{ah3_equn4}
I_G=k\int\Sigma_{p=0}^k{\alpha_pL^{(p)}} ,
\end{equation}
where $\alpha_p$-s are arbitrary (positive) coupling constants and $L^{(p)}$-s are the $p^{th}$ order dimensionally continued terms in Lagrangian and is given by  

\begin{equation}\label{ah3_equn5} 
L^{(p)}=\epsilon_{a_1...a_d}R^{a_1a_2}...R^{a_{{2p}-1}a_{2p}}e^{a_{2p+1}}...e^{a_d}~`,~~~p=0,1,...,k
\end{equation} 

In Lovelock model, the maximum order of terms $k$ in the action is related to the number of dimensions of the space-time as $k=[\frac{d-1}{2}]$, where $[x]$ represents the integer part of $x$. The metric which is derived from the action given in $\it eqn $ (\ref{ah3_equn4}), describes the spherically symmetric BH space-time with the choice of the resulting set of coupling constants labeled by the order $k$ and the gravitational constant $G_k $ as:
\begin{eqnarray}\label{ah3_equn6}
\alpha_p= \left\{\begin{array}{lr}
          \frac{R^{2(p-k)}}{d-2p}\left[k~~p\right]^T , & \text{for } p\leq k\\
        0, & \text{for } p>k\\
            \end{array}\right.
\end{eqnarray}\\
 
where $1\leq k\leq\left[ \frac{d-1}{2}\right]$. The resulting field equations would be of the form:
\begin{equation}\label{ah3_equn7}
\epsilon_{ba_1...a_{d-1}}\bar{R}^{a_1a_2}...\bar{R}^{a_{2k-1}}...\epsilon^{a_{d-1}}=0 ~~and
\end{equation}  
\begin{equation}\label{ah3_equn8}
\epsilon_{aba_3...a_d}\bar{R}^{a_3 a_4}...\bar{R}^{a_{2k-1}a_{2k}}T^{a_{2k+1}}...\epsilon^{a_{d-1}}=0.
\end{equation}  
Here we use the identity $\bar{R^{ab}}=R^{ab}+\frac{1}{R^2}e^{a+b}.$\\ 
The action given in $eqn.$ (\ref{ah3_equn4}) admits a static and spherically solution to $eqn.$ (\ref{ah3_equn7}) with the space-time metric, written in Schwarzchild like coordinates in the form given by:
\begin{equation}\label{ah3_equn9}
ds^2=f(r)dt^2+\frac{dr^2}{f(r)}+r^2d\Omega_{d-2}^2,
\end{equation}
where 
\begin{equation}\label{ah3_equn10}
f(r)=1+\frac{r^2}{R^2}-\sigma\left(\frac{C_1}{r^{d-2k-1}}\right)^{1/k}.
\end{equation}
The term $\Omega_{d-2}$ signifies the volume of the $(d-2)$ dimensional spherically symmetric tangent space and $R$ AdS radius. Here we take $\sigma$ as unity. The integrating constant $C_1$ is written as:
\begin{equation}\label{ah3_equn11}
C_1=2G_k(M+C_0),
\end{equation}
where $M$ denotes the mass of BH. The constant $C_0$ is chosen so that the horizon shrinks to a point for $M\rightarrow 0$, as:
\begin{equation}\label{ah3_equn12}
C_0=\frac{1}{2G_k}\delta_{d-2k,1}.
\end{equation}   
The interesting fact of $\it eqn$ (\ref{ah3_equn10}) is that the exponent of $\frac{1}{r}$ is proportional to $(d-2k-1)$. For even dimensions, this term has unit value and for odd dimensions it is zero, since $k=\left[\frac{d-1}{2}\right]$. The $(d-2k-1)=0$ cases correspond to Chern-Simpsons theories that is different form AdS \cite{Crisotomo 2015} due to presence of vacuum and the solution for condition $(d-2k-1)=1$ resemble the Schwarzchild-AdS solution.

The Hawking temperature of BH, related with the surface gravity on the horizon $r=r_h$ is given by \cite{Gen 2015}
\begin{equation}\label{ah3_equn13}
T_h = \frac{f'(r_h)}{4\pi}=\frac{r_h}{2\pi R^2}-\frac{1}{4\pi k}\left(2k+1-d\right)r^{2k-d}C_1\left(r_h^{(2k+1-d)}C_1\right)^{\frac{1}{k}-1}.
\end{equation}
The first order correction to entropy of BH is proportional to $\ln\left(C_V T^2\right)$, where $C_V=\frac{1}{T^2}\left(\frac{\partial^2 S}{\partial \beta^2}\right)_{\beta=\beta_0}$ \cite{Sommerfield1956}. Following some literatures as \cite{Pradhan2016, Das2001} we can compute $\frac{1}{T^2}\left(\frac{\partial^2 S}{\partial \beta^2}\right)_{\beta=\beta_0}=S_0$. Therefore, the correction term might be written as proportional to $\ln \left(S_0T^2\right)$. So the corrected entropy may be written as $$S=S_0-\frac{1}{2}\ln(S_0T_h^2).$$ We multiply the correction term with an arbitrary correction parameter or correction coefficient $\alpha$ which can regulate the effect of the correction term ($-1\leq \alpha \leq 1$). $\alpha$ should have dimension of length. So the modified entropy is now given by \cite{Das 2002}
\begin{equation}\label{ah3_equn14}
S=S_0-\frac{\alpha}{2}\ln(S_0T_h^2),
\end{equation}
where $S_0$ is the zeroth entropy and is given by \cite{XU 2015}
\begin{equation}\label{ah3_equn15}
S_0= \pi r_h^2~~.
\end{equation}
Using the $eqn.s $ (\ref{ah3_equn13}), (\ref{ah3_equn14}) and (\ref{ah3_equn15}) we obtain the corrected entropy as:
\begin{equation}\label{ah3_equn16}
S=\pi r_h^2-\frac{\alpha}{2}\ln \left[\frac{r_h^2}{16\pi}\bigg\{\frac{2r_h}{R^2}-\frac{1}{k}\left(2k+1-d\right)r_h^{2k-d} C_1\left(r_h^{2k+1-d}C_1\right)^{\frac{1}{k}-1}\bigg\}^2\right].
\end{equation} 
\begin{figure}[h!]
\begin{center}
~~~~~~~Fig.-1 ~~~~~~~~~~~~~~~~~~~\\
\includegraphics[scale=.8]{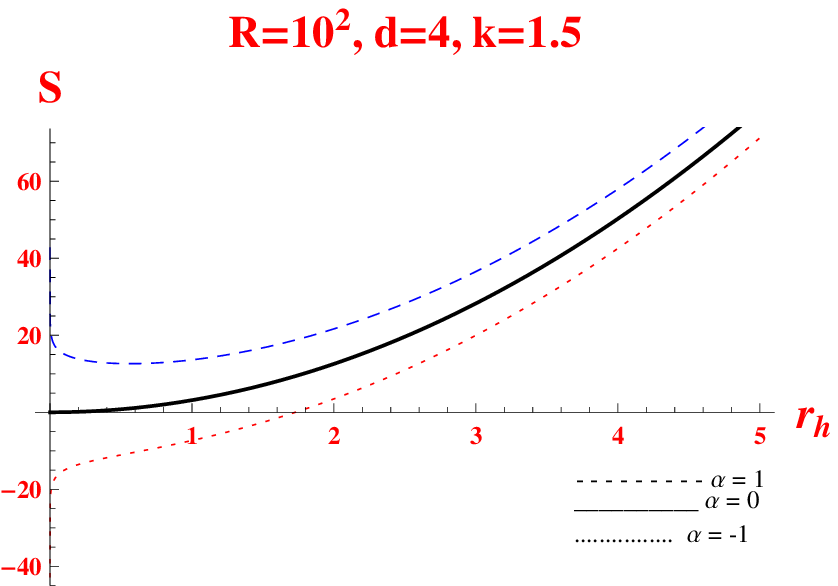}\\

Fig.-1 represents the variation of $ S $ with respect to horizon radius $ r_h $ keeping $ R $, $ d $ and hence $ k $ fixed with varying $ \alpha $ only. \\
\end{center} 
\end{figure}
We have plotted the variations of entropy $S$ with respect to horizon radius $r_h$, keeping $R$, $d$ and $k$ fixed with the variation of correction coefficient $\alpha$ in Fig.-1. Here we have considered 4-dimensional space-time. For low $r_h$ the values of entropies for different $\alpha$ values are almost same. But as we increase $r_h$, the BH grows bigger. This causes the area of event horizon to grow bigger and hence the entropy increases. For this particular type of BHs we observe that the entropy increases almost exponentially. The natures of curves for $\alpha=1$, $\alpha=0$ and $\alpha=-1$ curves are almost same for high event horizon region. But for same $r_h$, it is observed that the $S$ for  $\alpha=1>$ $S$ for  $\alpha=0>$ $S$ for  $\alpha=-1$.
  
The Helmholtz Free energy, related to entropy and temperature is expressed as:
\begin{equation}\label{ah3_equn17}
F=-\int {S dT_h}
\end{equation} 
and we find the equation (\ref{ah3_equn29}) for $F$ which is given in appendix I for huge mathematical structure.
\begin{figure}[h!]
\begin{center}
~~~~~~~Fig.-2 ~~~~~~~~~~~~~~~~~~~\\
\includegraphics[scale=.8]{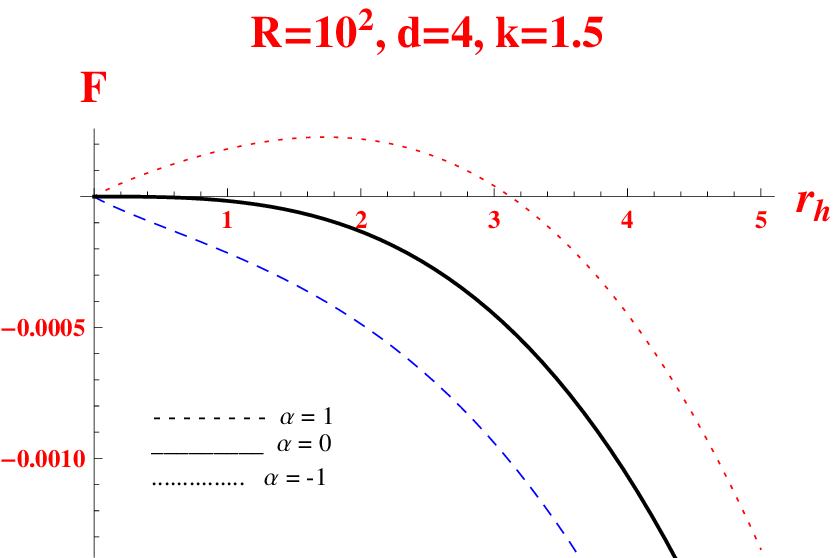}

Fig.-2 represents the variation of $ F $ with respect to horizon radius $ r_h $ keeping $ R $, $ d $ and hence $ k $ are fixed with varying $ \alpha $ only. 
\end{center} 
\end{figure}

Fig.-2 shows the variation Helmholtz free energy $F$ with $r_h$. For $\alpha=-1$, we see $F$ is increasing for low $r_h$. Then after reaching a local maxima  it starts to decrease. For $\alpha=0$ and $\alpha=1$, $F$ is decreasing always. But for same $r_h$ we see $F$ (for $\alpha=1$)$>F(for~\alpha=1)$. For $\alpha=-1$ only, if radius of event horizon $r_h$ is low we observe $F$ to be positive. Otherwise it possesses a negative value. Negative $F$ signifies the fact that we will not be able to extract any useful work out of the concerned BH. In $\alpha=-1$ case, as $F$ changes the sign inbetween (say at $r_h=r_{hcrit_1}$), it signifies a phase transition.
  
The internal energy is to be calculated from the well-known thermodynamic relation, given as: \begin{equation}\label{ah3_equn18}
U=F + T_h S
\end{equation} and we obtain the equation (\ref{ah3_equn30}) of appendix-I for the internal energy.
\begin{figure}[h!]
\begin{center}
~~~~~~~Fig.-3 ~~~~~~~~~~~~~~~~~~~\\
\includegraphics[scale=.8]{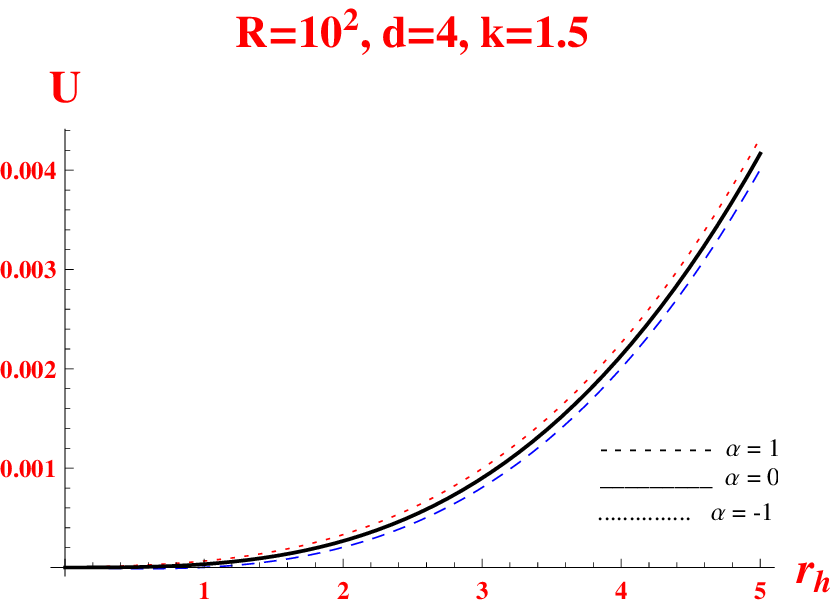}

Fig.-3 represents the variation of $ U $ with respect to horizon radius $ r_h $ keeping $ R $, $ d $ and hence $ k $ are fixed with varying $ \alpha $ only. 
\end{center} 
\end{figure}\\
The thermodynamic volume of the BH is given by 
\begin{equation}\label{ah3_equn19}
V=\frac{4}{3}\pi r_h^3
\end{equation}
and the modified pressure of the black BH due to the thermal fluctuation can be derived from the relation,
\begin{equation}\label{ah3_equn20}
P=-\left(\frac{\partial F}{\partial V}\right)_T=-\left(\frac{\frac{\partial F}{\partial r_h}}{\frac{\partial V}{\partial r_h}}\right)_T
\end{equation}
which gives $eqn.$ (\ref{ah3_equn31}) of appendix-I.
It is evident from $eqn.$ (\ref{ah3_equn20}) that when $r_h$ is increasing function of $\alpha$ for small radius, pressure is decreasing function of $r_h$ as expected. We also notice that for large event horizon radius the logarithmic correction does not play an important role on the pressure of the BH.\\

Another thermodynamic parameter `enthalpy' can be derived from the relation given by
\begin{equation}\label{ah3_equn21}
H=U+PV
\end{equation}
and we have the equation (\ref{ah3_equ32}) of appendix I.
which is also decreasing function of $\alpha$ as well.\\
\begin{figure}[h!]
\begin{center}
~~~~~~~Fig.-4 ~~~~~~~~~~~~~~~~~~~\\
\includegraphics[scale=.8]{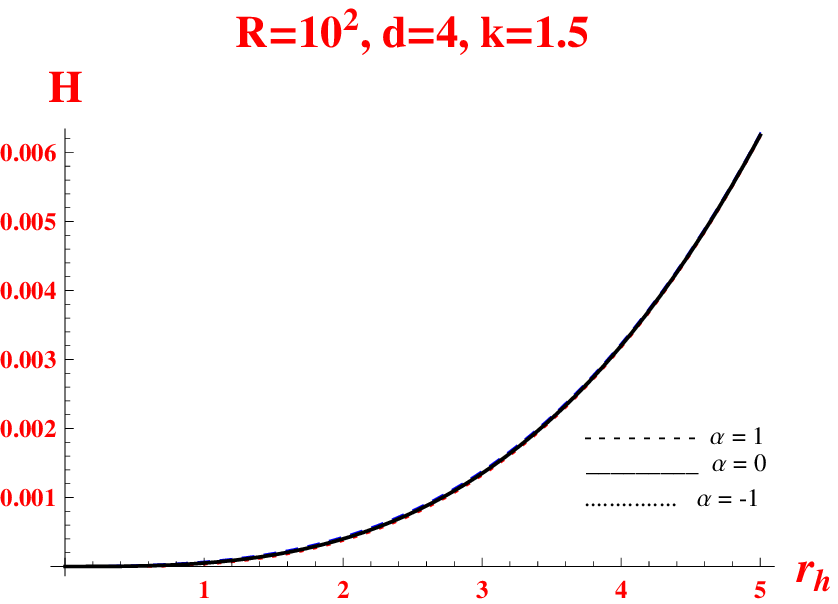}\\

Fig.-4 represents the variation of $ H $ with respect to horizon radius $ r_h $ keeping $ R $, $ d $ and hence $ k $ are fixed with varying $ \alpha $ only. 
\end{center} 
\end{figure}

The Gibbs free energy can be obtained from the following thermodynamic relation as:
\begin{equation}\label{ah3_equn22}
G=H - T_h S=F+PV,
\end{equation} and we get the equation (\ref{ah3_equn33}) of appendix-I.

We find that Gibbs free energy is decreasing function of $\alpha$ like all other thermodynamic parameters already have been studied.\\
\begin{figure}[h!]
\begin{center}
~~~~~~~Fig.-5 ~~~~~~~~~~~~~~~~~~~\\
\includegraphics[scale=.8]{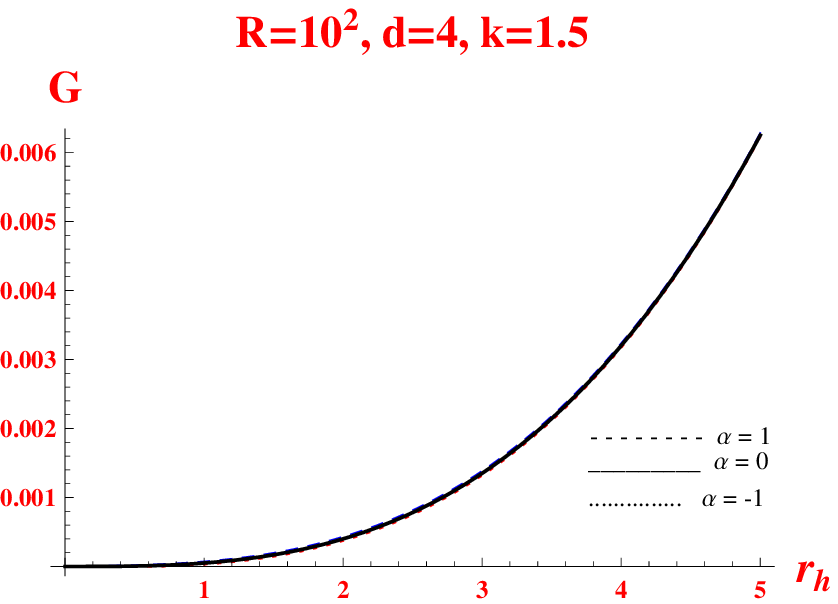}\\

Fig.-5 represents the variation of $ G $ with respect to horizon radius $ r_h $ keeping $ R $, $ d $ and hence $ k $ are fixed with varying $ \alpha $ only. 
\end{center} 
\end{figure}

\section{P-V Criticality and Stability:}
The critical point is obtained as the inflection point in the P-V diagram via the following equations:
\begin{equation}\label{ah3_equn23}
\left.\frac{\partial P}{\partial V}\right| _{T = T_c}=0
\end{equation}\\
and
\begin{equation}\label{ah3_equn24}
\left.\frac{\partial^2 P}{\partial V^2}\right| _{T = T_c}=0.
\end{equation}\\
In 4-dimensional space-time, at $\alpha=1$ limit, one can find the `critical pressure' $P_c$, `critical volume' $V_c$ and `critical temperature' $T_c$ by solving $eqns$ (\ref{ah3_equn23}), (\ref{ah3_equn20}), (\ref{ah3_equn19}) and (\ref{ah3_equn13}) as:
$$P_c=\frac{1}{8\pi R^2}\left(\frac{1}{2e\pi^\frac{3}{2}R^2}-1\right),$$
$$V_c=\frac{8\sqrt{2}}{3}e^{\frac{3}{2}}\pi^{\frac{7}{4}} R^3,$$
and 
\begin{equation}\label{ah3_equn25}
T_c=\frac{\sqrt{e}}{\sqrt{2}\pi^{\frac{3}{4}}R}.
\end{equation}\\
Again, by using $eqn$ (\ref{ah3_equn24}) for the sameone can get   
$$P_c=e^\frac{5}{3}\sqrt{\pi}-\frac{5}{6\pi R^2}, $$
$$V_c=\frac{8\sqrt{2}}{3}e^{\frac{5}{2}}\pi^{\frac{7}{4}} R^3,$$
and 
\begin{equation}\label{ah3_equn26}
T_c=\frac{e^\frac{5}{6}}{\sqrt{2}\pi^{\frac{3}{4}}R}.
\end{equation}\\
It is clear from the $eqn.s$ (\ref{ah3_equn25}) and (\ref{ah3_equn26}) that they do not satisfy simultaneously, which means that there is no critical point without thermal fluctuations. So to have the critical points we should consider the effect of thermal fluctuations.
\begin{figure}[h!]
\begin{center}
~~~~~~~Fig.-6a ~~~~~~~~~~~~~~~~~~~\\
\includegraphics[scale=.8]{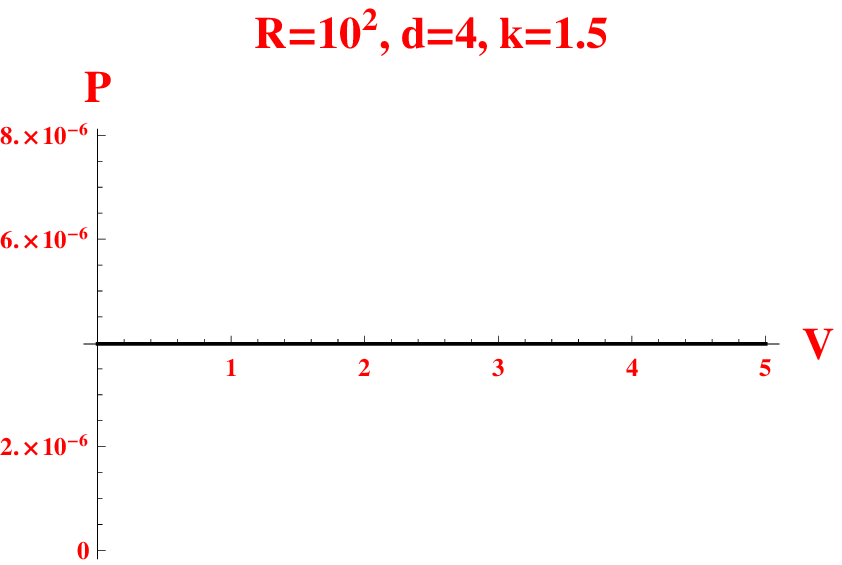}

Fig.-6a represents the variation of $ P $  with respect to the thermodynamic volume $ V $ for $\alpha=0$.
\end{center} 
\end{figure}
\begin{figure}[h!]
\begin{center}
~~~~~~~~~~~~~~~~~~~~~~Fig.-6b ~~~~~~~~~~~~~~~~~~~~~~~~~~~~~~~~~~~~Fig.-6c~~~~~~~~~~~~~~~~~~~~~~\\
\includegraphics[scale=.8]{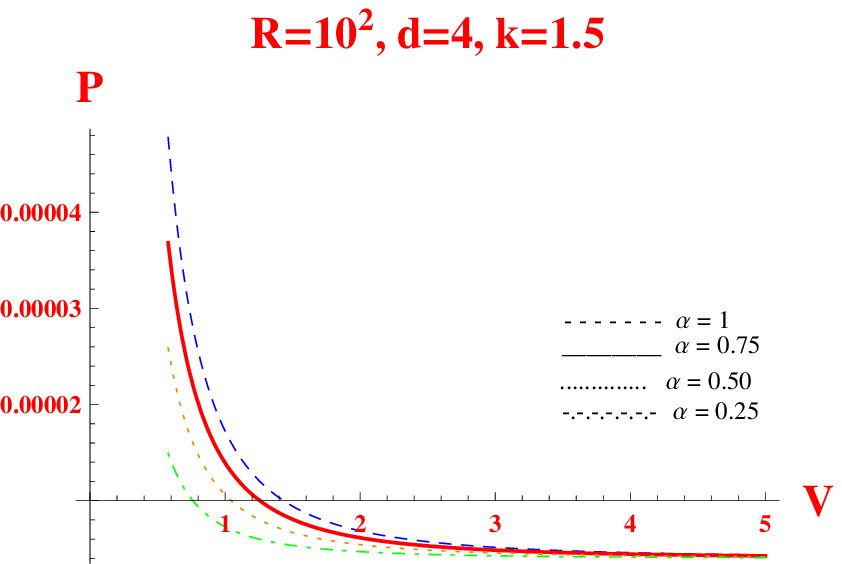}
\includegraphics[scale=.8]{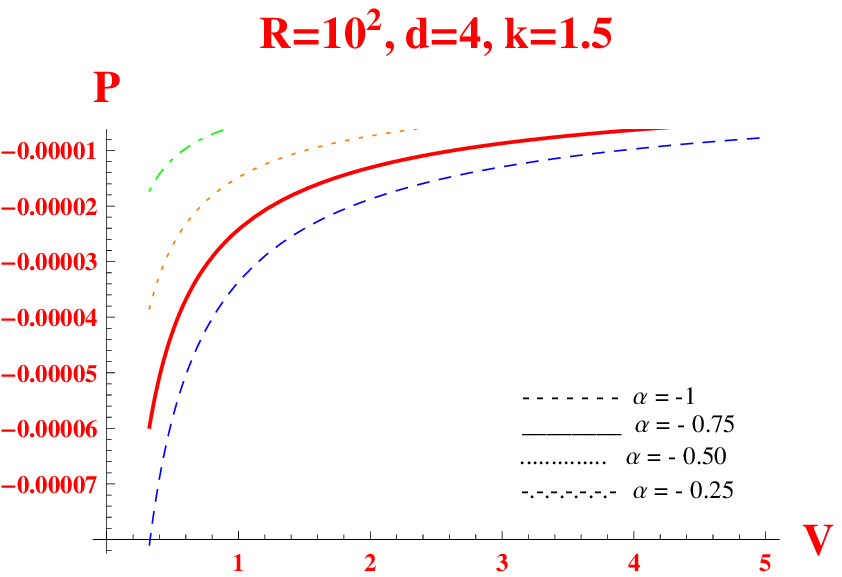}

Fig.-6b represents the variation of $ P $  with respect to the thermodynamic volume $ V $for $\alpha>0$.\\
Fig.-6c represents the variation of $ P $  with respect to the thermodynamic volume $ V $for $\alpha<0$.
\end{center} 
\end{figure}\\
Fig.-6a depicts the variation of $P$ with respect to $V$ for $\alpha=0$. We observe that the curve is parallel to the $V$ axis. So this signifies a constant pressure system. The increment of $V$ is caused by the increment in temperature. Fig,-6b says about the variation of $P$ with $V$ for positive $\alpha$-s. The curves make after the simple Boyle's law case. For a particular $V$, $P$ for $\alpha=1>$ $P$ for $\alpha=0.75>$ $P$ for $\alpha=0.50>$ $P$ for $\alpha=0.25$. So comparing with simple Boyle's law case i.e., the curves are rectangular hyperbolas: For low volume we observe high pressure and for high volume the pressure reduces down. We can say $\alpha$  is analogous to the temperature. Higher is $\alpha$ (or temperature), higher will be the value of pressure at a particular volume. For $\alpha<0$ cases we see a sets of rectangular hyperbolas but the pressure is found to be negative. This situation, however can be taken as an unphysical one or so exotic incident might be pointed out.
   
Now we study the stability of the BH by employing the specific heat which is given by,
\begin{equation}\label{ah3_equn27}
C=T\left(\frac{dS}{dT}\right)=T\left(\frac{\frac{dS}{dr_h}}{\frac{dT}{dr_h}}\right) ,
\end{equation}
and examine it in different regime. $eqn$ (\ref{ah3_equn27}) yields the specific heat of the BH as:
\begin{equation}\label{ah3_equn28}
C=\frac{\{\frac{2r_h}{R^2}-\frac{1}{k}(2k-d+1)r_h^{2k-d}C_1(r_h^{2k-d+1}C_1)^{\frac{1}{k}-1}\}\left(2\pi r_h -\frac{\alpha}{r_h}\right)}{\frac{2}{R^2}-\frac{1}{k}\left(\frac{1}{k}-1\right)(2k+1-d)^2r_h^{4k-2d}C_1^2(r_h^{2k+1-d}C_1)^{\frac{1}{k}-2}-(2k-d)(2k+1-d)r_h^{2k-1-d}C_1(r_h^{2k+1-d}C_1)^{\frac{1}{k}-1}}-\alpha 
\end{equation}

\begin{figure}[h!]
\begin{center}
~~~~~~~Fig.-7 ~~~~~~~~~~~~~~~~~~~\\
\includegraphics[scale=.8]{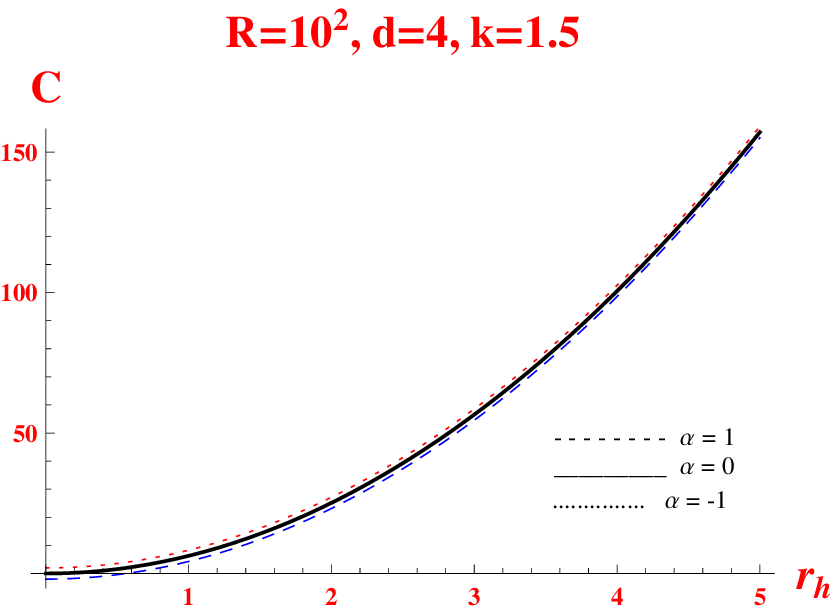}

Fig.-7 represents the variation of $ C $ with respect to horizon radius $ r_h $ keeping $ R $, $ d $ and hence $ k $ are fixed with varying $ \alpha $ only. 
\end{center} 
\end{figure}
In $Fig.-7$ we have plotted $C$ vs $r_h$ for $\alpha=-1, 0, 1$, we observe that the curves of heat capacity $C$ are always positive and increasing function. Positive heat capacity signifies a stable BH.\\
\section{Geometrothermodynamic Studies}
We calculate the Weinhold metric for the thermodynamic space of the concerned BH. Due to huge structure we in corporate it into equation (\ref{ah3_equn34}) of Appendix-I. 
The corresponding Ruppeiner metric can be easily calculated by dividing it by $T_h$. We calculate the Ricci scalar for this metric and plot this Ricci scalar with respect to $r_h$ and $R$ in the Fig.-8.
\begin{figure}[h!]
\begin{center}
~~~~~~~Fig.-8 ~~~~~~~~~~~~~~~~~~~\\
\includegraphics[scale=.8]{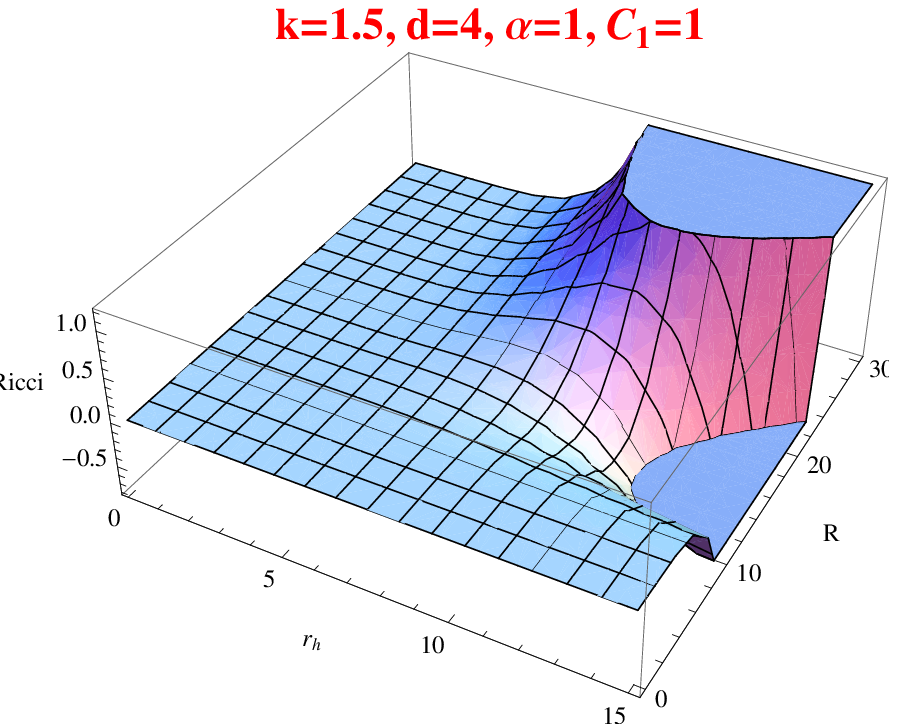}~~\\

Fig.-8 represents the variation of $ Ricci~Scalar $ of Ruppeiner metric with respect to horizon radius $ r_h $ keeping $ R $, $ d $ and $ k $ are fixed with  $ \alpha=1$. 
\end{center} 
\end{figure}
It is observed that if $R$ is low, the Ricci scalar stays almost constant with respect to $r_h$. If we increase $R$, after a particular value of $R=R_{crit_1}$, it is observed that Ricci scalar decreases steeply as we increase $r_h$. Again , after crossing another value of $R+R_{crit_2}$, the value of Ricci increases abruptly if we increase $r_h$. This peculiar behavior indicates that after we take more than a particular value of $R$, we can observe that the curve of Ricci attains a local maxima before which it increases and after which it decreases down. With $R$, this local maxima increseas as well. But no clear divergence or a jump discontinuity is observed. This speculates the absence of any phase transition.
\section{Conclusions}
In our study, we have investigated the effect of thermal fluctuations on the thermodynamics of a Lovelock-AdS BH. Here, we have considered the BH in 4-dimensional space-time. We have computed the thermodynamic potentials like first order logarithamic correction term in entropy then Helmholtz free energy, Enthalpy and Gibbs free energy of the said BH and we observe that all the thermodynamic potentials are decreasing functions of correction coefficient $\alpha$.

We have plotted the BH entropy $S$ and the Helmholtz free energy $ F $  with respect to horizon radius $r_h$, keeping $R$, $d$ and $k$ are fixed with the variation of correction coefficient $\alpha$ in Fig.-1 and Fig.-2. We have found that for low $r_h$ the values of entropies for different values of $\alpha$ are almost same. But due to increment of $r_h$ the BH grows bigger. This causes the area of event horizon to grow bigger and hence the entropy increases in Fig.-1 and in Fig.-2 we have noticed that the critical values of horizon radius does not depend upon the AdS radius $R$ it depends only correction coefficient $\alpha$ only.

Further we have computed the $P-V$ criticality of this BH. We have calculated the critical volume $V_c$, critical pressure $P_c$ and critical temperature $T_c$ by using different equations and show that there is no critical point for this BH in absence of logarithmic correction. So to have the critical points we should consider the effect of thermal fluctuations i.e., logarithmic correction. Finally we have examined the thermal stability of the BH by employing the specific heat $C$ in different regime in $Fig.-7$ and we have observed that the curves of heat capacity $C$ are always positive and increasing function. Positive heat capacity signifies a stable BH.

When we plot $P-V$ diagrams (Fig.-6a-c) for $\alpha<0$, $\alpha=0$ and $\alpha>0$ we find the BHs to behave as a constant pressure system for $\alpha=0$. It has no physical existence for $\alpha<0$. But for $\alpha>0$, the $P-V$ diagram is analogous to Boyle's law which means $\alpha$ can have only positive values, i.e., the correction parameter or correction coefficient which we have introduced must be positive. 

In a nutshell, our work has shown that the thermodynamic behavior of Lovelock AdS black holes resembles that of the other AdS black holes. AdS radius is a crucial parameter to determine how large value the Ricci scalar of the Ruppeiner metric of such AdS black holes can attain. We did not find any phase transition depicted by the natures of the specific heat, free energies or the Ricci scalar of Ruppeiner metric. It is predicted that the value of the correction coefficient of the entropy preferably should be a positive parameter.
\vspace{.1 in}
{\bf Acknowledgment:}
This research is supported by the project grant of Goverment of West Bengal, Department of Higher Education, Science and Technology and Biotechnology (File no:- $ST/P/S\&T/16G$-$19/2017$). AH wishes to thank the Department of Mathematics, the University of Burdwan for the research facilities provided during the work. RB thanks IUCAA, Pune, India for providing Visiting Associateship.


\section{Appendix-I}
The Helmholtz Free Energy takes the form
$$F=\frac{1}{8k\pi r_hR^2}\Bigg(\alpha\ln \left[\frac{1}{16\pi}\bigg\{\frac{2r_h^2}{R^2}+\frac{1}{k}\left(2k-1+d\right)\left(r_h^{2k+1-d}C_1\right)^{\frac{1}{k}}\bigg\}^2\right]\Bigg[\left(d-1\right)R^2 \left(r_h^{2k+1-d}C_1\right)^{\frac{1}{k}}+2k\bigg\{r_h^2-R^2 $$
$$\times\left(r_h^{2k+1-d}C_1\right)\bigg\}^{\frac{1}{k}}\Bigg]\Bigg)-\frac{\left(d-2k-1\right)R^2\left[(k+1-d)^2 \pi r_h^2+\{1+d^2+5k+6k^2-d(2+5k)\}\alpha\right]\left(r_h^{1+2k-d}C_1\right)^{\frac{1}{k}}}{4k\pi r_hR^2(d-3k-1)(d-k-1)}$$
\begin{equation}\label{ah3_equn29}
-\frac{r_h^3}{6R^2}-\frac{r_h\alpha}{\pi R^2}.
\end{equation}
The internal energy is given by
$$U=\frac{1}{8k\pi r_hR^2}\Bigg(\alpha\ln \left[\frac{1}{16\pi}\bigg\{\frac{2r_h^2}{R^2}+\frac{1}{k}\left(2k-1+d\right)\left(r_h^{2k+1-d}C_1\right)^{\frac{1}{k}}\bigg\}^2\right]\Bigg[\left(d-1\right)R^2 \left(r_h^{2k+1-d}C_1\right)^{\frac{1}{k}}+2k\bigg\{r_h^2-R^2 $$

$$\times\left(r_h^{2k+1-d}C_1\right)\bigg\}^{\frac{1}{k}}\Bigg]\Bigg)-\frac{\left(d-2k-1\right)R^2\left[(k+1-d)^2 \pi r_h^2+\{1+d^2+5k+6k^2-d(2+5k)\}\alpha\right]\left(r_h^{1+2k-d}C_1\right)^{\frac{1}{k}}}{4k\pi r_hR^2(d-3k-1)(d-k-1)}$$

$$-\frac{r_h^3}{6R^2}-\frac{r_h\alpha}{\pi R^2}+\frac{1}{4\pi}\left(\pi r_h^2-\frac{\alpha}{2}\ln \left[\frac{r_h^2}{16\pi}\bigg\{\frac{2r_h}{R^2}-\frac{1}{k}\left(2k+1-d\right)r_h^{2k-d} C_1\left(r_h^{2k+1-d}C_1\right)^{\frac{1}{k}-1}\bigg\}^2\right]\right)$$ 

\begin{equation}\label{ah3_equn30}
\times\left[\frac{2r_h}{R^2}-\frac{1}{k}\left(2k+1-d\right)r_h^{2k-d} C_1\left(r_h^{2k+1-d}C_1\right)^{\frac{1}{k}-1}\right].
\end{equation} 
Modified pressure of the black hole can be given as
$$P=\frac{\left(d-2k-1\right)R^2\left[(k+1-d)^2 \pi r_h^2+\{1+d^2+5k+6k^2-d(2+5k)\}\alpha\right]\left(r_h^{1+2k-d}C_1\right)^{\frac{1}{k}}}{16k\pi^2 r_h^2R^2(d-3k-1)(d-k-1)}$$
$$-\frac{r_h}{24\pi R^2}-\frac{\alpha}{4\pi^2r_h R^2}-\frac{1}{32k\pi^2 r_h^3R^2}\times\Bigg(\alpha\ln \left[\frac{1}{16\pi}\bigg\{\frac{2r_h^2}{R^2}+\frac{1}{k}\left(2k-1+d\right)\left(r_h^{2k+1-d}C_1\right)^{\frac{1}{k}}\bigg\}^2\right]$$

$$\times\Bigg[\left(d-1\right)R^2 \left(r_h^{2k+1-d}C_1\right)^{\frac{1}{k}}+2k\{r_h^2-R^2\left(r_h^{2k+1-d}C_1\right)\}^{\frac{1}{k}}\Bigg]\Bigg)$$

$$+\frac{(d-2k-1)(2k+1-d)r_h^{2k-d}R^2[(k+1-d)^2\pi r_h^2+\{1+d^2+5k+6k^2-d(2+5k)\}\alpha]C_1\left(r_h^{2k+1-d}C_1\right)^{\frac{1}{k}-1}}{16k^2\pi^2 r_h^2R^2(d-3k-1)(d-k-1)}$$

$$-\frac{1}{6\pi R^2}+\frac{\alpha}{2\pi^2 r_h R^2}+\frac{\alpha}{32k\pi^2 r_h^3R^2}\ln \left[\frac{1}{16\pi}\bigg\{\frac{2r_h^2}{R^2}+\frac{1}{k}\left(2k-1+d\right)\left(r_h^{2k+1-d}C_1\right)^{\frac{1}{k}}\bigg\}^2\right]\times$$

$$\left[\frac{1}{k}(d-1)(2k+1-d)r_h^{2k-d}R^2C_1\left(r_h^{2k+1-d}C_1\right)^{\frac{1}{k}-1}+4kr_h-2(2k+1-d)r_h^{2k-d}R^2C_1\left(r_h^{2k+1-d}C_1\right)^{\frac{1}{k}-1}\right]-$$
\begin{equation}\label{ah3_equn31}
\frac{\alpha\{\frac{4r_h}{R^2}+\frac{1}{k^2}(d-2k-1)(2k+1-d)r_h^{2k-d}C_1\left(r_h^{2k+1-d}C_1\right)^{\frac{1}{k}-1}\}[(d-1)R^2(r_h^{2k+1-dC_1})^\frac{1}{k}+2k\{r_h^2-R^2(r_h^{2k+1-d}C_1)^\frac{1}{k}\}]}{16k\pi^2 r_h^3R^2\{\frac{2r_h^2}{R^2}+\frac{1}{k}(d-2k-1)(r_h^{2k+1-d}C_1)^\frac{1}{k}\}}.
\end{equation}
The enthalpy reads as $$H=\frac{1}{8k\pi r_hR^2}\Bigg(\alpha\ln \left[\frac{1}{16\pi}\bigg\{\frac{2r_h^2}{R^2}+\frac{1}{k}\left(2k-1+d\right)\left(r_h^{2k+1-d}C_1\right)^{\frac{1}{k}}\bigg\}^2\right]\Bigg[\left(d-1\right)R^2 \left(r_h^{2k+1-d}C_1\right)^{\frac{1}{k}}+2k\bigg\{r_h^2-R^2 $$

$$\times\left(r_h^{2k+1-d}C_1\right)\bigg\}^{\frac{1}{k}}\Bigg]\Bigg)-\frac{\left(d-2k-1\right)R^2\left[(k+1-d)^2 \pi r_h^2+\{1+d^2+5k+6k^2-d(2+5k)\}\alpha\right]\left(r_h^{1+2k-d}C_1\right)^{\frac{1}{k}}}{4k\pi r_hR^2(d-3k-1)(d-k-1)}$$

$$+\frac{r_h^3}{18R^2}-\frac{r_h\alpha}{\pi R^2}+\frac{1}{4\pi}\left(\pi r_h^2-\frac{\alpha}{2}\ln \left[\frac{r_h^2}{16\pi}\bigg\{\frac{2r_h}{R^2}-\frac{1}{k}\left(2k+1-d\right)r_h^{2k-d} C_1\left(r_h^{2k+1-d}C_1\right)^{\frac{1}{k}-1}\bigg\}^2\right]\right)$$ 

$$\times\left[\frac{2r_h}{R^2}-\frac{1}{k}\left(2k+1-d\right)r_h^{2k-d} C_1\left(r_h^{2k+1-d}C_1\right)^{\frac{1}{k}-1}\right]-\frac{r_h^4}{18R^2}+\frac{r_h^2\alpha}{3\pi R^2}$$

$$+\frac{r_h\left(d-2k-1\right)R^2\left[(k+1-d)^2 \pi r_h^2+\{1+d^2+5k+6k^2-d(2+5k)\}\alpha\right]\left(r_h^{1+2k-d}C_1\right)^{\frac{1}{k}}}{12k\pi R^2(d-3k-1)(d-k-1)}-\frac{1}{24k\pi r_hR^2}\Bigg(\alpha\ln$$

$$ \times\left[\frac{1}{16\pi}\bigg\{\frac{2r_h^2}{R^2}+\frac{1}{k}\left(2k-1+d\right)\left(r_h^{2k+1-d}C_1\right)^{\frac{1}{k}}\bigg\}^2\right]\left[\left(d-1\right)R^2 \left(r_h^{2k+1-d}C_1\right)^{\frac{1}{k}}+2k\bigg\{r_h^2-R^2\left(r_h^{2k+1-d}C_1\right)\bigg\}^{\frac{1}{k}}\right]\Bigg)$$

$$+\frac{(d-2k-1)(2k+1-d)r_h^{2k+1-d}R^2[(k+1-d)^2\pi r_h^2+\{1+d^2+5k+6k^2-d(2+5k)\}\alpha]C_1\left(r_h^{2k+1-d}C_1\right)^{\frac{1}{k}-1}}{12k^2\pi R^2(d-3k-1)(d-k-1)}-$$

$$\frac{\alpha}{24k\pi R^2}\ln \left[\frac{1}{16\pi}\bigg\{\frac{2r_h^2}{R^2}+\frac{1}{k}\left(2k-1+d\right)\left(r_h^{2k+1-d}C_1\right)^{\frac{1}{k}}\bigg\}^2\right]\times$$

$$\left[\frac{1}{k}(d-1)(2k+1-d)r_h^{2k-d}R^2C_1\left(r_h^{2k+1-d}C_1\right)^{\frac{1}{k}-1}+4kr_h-2(2k+1-d)r_h^{2k-d}R^2C_1\left(r_h^{2k+1-d}C_1\right)^{\frac{1}{k}-1}\right]-$$
\begin{equation}\label{ah3_equ32}
\frac{\alpha\{\frac{4r_h}{R^2}+\frac{1}{k^2}(d-2k-1)(2k+1-d)r_h^{2k-d}C_1\left(r_h^{2k+1-d}C_1\right)^{\frac{1}{k}-1}\}[(d-1)R^2(r_h^{2k+1-dC_1})^\frac{1}{k}+2k\{r_h^2-R^2(r_h^{2k+1-d}C_1)^\frac{1}{k}\}]}{12k\pi R^2\{\frac{2r_h^2}{R^2}+\frac{1}{k}(d-2k-1)(r_h^{2k+1-d}C_1)^\frac{1}{k}\}}.
\end{equation}\\
The Gibbs free energy has the form
$$G=\frac{1}{8k\pi r_hR^2}\Bigg(\alpha\ln \left[\frac{1}{16\pi}\bigg\{\frac{2r_h^2}{R^2}+\frac{1}{k}\left(2k-1+d\right)\left(r_h^{2k+1-d}C_1\right)^{\frac{1}{k}}\bigg\}^2\right]\Bigg[\left(d-1\right)R^2 \left(r_h^{2k+1-d}C_1\right)^{\frac{1}{k}}+2k\bigg\{r_h^2-R^2 $$
$$\times\left(r_h^{2k+1-d}C_1\right)\bigg\}^{\frac{1}{k}}\Bigg]\Bigg)-\frac{\left(d-2k-1\right)R^2\left[(k+1-d)^2 \pi r_h^2+\{1+d^2+5k+6k^2-d(2+5k)\}\alpha\right]\left(r_h^{1+2k-d}C_1\right)^{\frac{1}{k}}}{4k\pi r_hR^2(d-3k-1)(d-k-1)}+\frac{r_h^3}{18R^2}$$

$$-\frac{r_h\alpha}{\pi R^2}+\frac{r_h\left(d-2k-1\right)R^2\left[(k+1-d)^2 \pi r_h^2+\{1+d^2+5k+6k^2-d(2+5k)\}\alpha\right]\left(r_h^{1+2k-d}C_1\right)^{\frac{1}{k}}}{12k\pi R^2(d-3k-1)(d-k-1)}-\frac{r_h^4}{18R^2}+\frac{r_h^2\alpha}{3\pi R^2}-$$

$$\frac{1}{24k\pi r_hR^2}\left(\alpha\ln \left[\frac{1}{16\pi}\bigg\{\frac{2r_h^2}{R^2}+\frac{1}{k}\left(2k-1+d\right)\left(r_h^{2k+1-d}C_1\right)^{\frac{1}{k}}\bigg\}^2\right]\left[\left(d-1\right)R^2 \left(r_h^{2k+1-d}C_1\right)^{\frac{1}{k}}+2k\bigg\{r_h^2-R^2\left(r_h^{2k+1-d}C_1\right)\bigg\}^{\frac{1}{k}}\right]\right)+$$

$$+\frac{(d-2k-1)(2k+1-d)r_h^{2k+1-d}R^2[(k+1-d)^2\pi r_h^2+\{1+d^2+5k+6k^2-d(2+5k)\}\alpha]C_1\left(r_h^{2k+1-d}C_1\right)^{\frac{1}{k}-1}}{12k^2\pi R^2(d-3k-1)(d-k-1)}-$$

$$\frac{\alpha}{24k\pi R^2}\ln \left[\frac{1}{16\pi}\bigg\{\frac{2r_h^2}{R^2}+\frac{1}{k}\left(2k-1+d\right)\left(r_h^{2k+1-d}C_1\right)^{\frac{1}{k}}\bigg\}^2\right]\times$$

$$\left[\frac{1}{k}(d-1)(2k+1-d)r_h^{2k-d}R^2C_1\left(r_h^{2k+1-d}C_1\right)^{\frac{1}{k}-1}+4kr_h-2(2k+1-d)r_h^{2k-d}R^2C_1\left(r_h^{2k+1-d}C_1\right)^{\frac{1}{k}-1}\right]-$$
\begin{equation}\label{ah3_equn33}
\frac{\alpha\{\frac{4r_h}{R^2}+\frac{1}{k^2}(d-2k-1)(2k+1-d)r_h^{2k-d}C_1\left(r_h^{2k+1-d}C_1\right)^{\frac{1}{k}-1}\}[(d-1)R^2(r_h^{2k+1-dC_1})^\frac{1}{k}+2k\{r_h^2-R^2(r_h^{2k+1-d}C_1)^\frac{1}{k}\}]}{12k\pi R^2\{\frac{2r_h^2}{R^2}+\frac{1}{k}(d-2k-1)(r_h^{2k+1-d}C_1)^\frac{1}{k}\}}.
\end{equation}

The Weinhold Metric takes the form
\begin{equation}\label{ah3_equn34}
ds^2_W=\frac{kr_h^{d-2k+1}\left(1+\frac{r_h^2}{R^2}\right)^k}{G_kR^2\left(r_h^2+R^2\right)^2}\left[AdS^2+2BdSdR+CdR^2\right]
\end{equation}
where,
 $$A=\frac{r_h^2R^2}{2}\left[(d-2)(d-1)r_h^4+2\left\{2+d(d-2k-3)+5k\right\}r_h^2R^2+(d-2k-1)\left\{d-2(k+1)\right\}R^4\right]$$
 $$\times\left\{2kr_h^2+(d-2k-1)R^2(C_1r_h^{1-d+2k})^{\frac{1}{k}}\right\}^2$$
 $$\times\left[8k^3r_h^4(\pi r_h^2+\alpha)-2(d-2k-1)r_h^2R^2\left\{(d-1)\alpha+(d-1+k)-4k^2(\pi r_h^2+\alpha)\right\}(C_1r_h^{1-d+2k})^{\frac{1}{k}}\right.$$
 $$\left[1+(d-2k-1)^2R^4\left\{\alpha(1-d)+2k(\pi r_h^2+\alpha)\right\}(C_1r_h^{1-d+2k})^{\frac{1}{k}}]\right]^{-1}~~,$$
 $$B=2kR\left[-8k^3r_h^4\left\{(d-1)r_h^2+(d-2k+1)R^2\right\}(\pi r_h^2-\alpha)-2R^2(d-2k-1)\left\{4\pi k^2r_h^4[(d-1)r_h^2+(d-2k+1)R^2]+\right.\right.$$
 $$\left.[(d-1)\{(d-1)(1+k-4k^2)\}r_h^4+(d-2k-1)\{d(2+k)+k(3-4k)-2\}r_h^2R^2+(d-1)(d-2k-1)R^4]\alpha\right\}(C_1r_h^{1-d+2k})^{\frac{1}{k}}$$
 $$\left.-(d-2k-1)^2R^4\{(d-1)r_h^2+(d-2k+1)R^2\}\{2k(\pi r_h^2-\alpha)+(d-1)\alpha\}(C_1r_h^{1-d+2k})^{\frac{1}{k}}\right]$$
 $$\times [4k^2r_h^2(\pi r_h^2-\alpha)+(d-2k-1)R^2\{2k(\pi r_h^2-\alpha)+(d-1)\alpha\}(C_1r_h^{1-d+2k})^{\frac{1}{k}}]^{-2}$$
 and
 $$C=\{(2k+1)r_h^2+3R^2\}$$
 \end{document}